%% file: paper.tex
\documentclass{nature}

\usepackage{graphicx}  
\usepackage{braket}
\usepackage{color}
\usepackage{mathtools}
\usepackage{bm}
\usepackage{mathrsfs}
\usepackage[implicit=false]{hyperref}

\newcommand{\mbf}{\mathbf}
\newcommand{\mrm}{\mathrm}

\newcommand\blfootnote[1]{%
  \begingroup
  \renewcommand\thefootnote{}\footnote{#1}%
  \addtocounter{footnote}{-1}%
  \endgroup
}

\title{Quantifying entanglement in a 68-billion dimensional quantum state space}

\author{James Schneeloch$^1$, Christopher C. Tison$^{1,2,3}$, Michael L. Fanto$^{1,4}$, Paul M. Alsing$^1$ \& Gregory A. Howland$^{1,4,^*}$}

\begin{document}

\maketitle

\begin{affiliations}
 \item Air Force Research Laboratory, Information Directorate, Rome, NY 13441 USA
 \item Department of Physics, Florida Atlantic University, Boca Raton, FL 33431 USA
 \item Quanterion Solutions Incorporated, Utica, NY 13502 USA
 \item Rochester Institute of Technology, Rochester, NY 14623 USA
\end{affiliations}

\begin{abstract}
\blfootnote{This is a post-peer review, pre-copyedit version of an article published in Nature Communications. The final authenticated version is available online at: \url{http://dx.doi.org/10.1038/s41467-019-10810-z}}
Entanglement is the powerful and enigmatic resource central to quantum information processing, which promises capabilities in computing, simulation, secure communication, and metrology beyond what is possible for classical devices. Exactly quantifying the entanglement of an unknown system requires completely determining its quantum state, a task which demands an intractable number of measurements even for modestly-sized systems. Here we demonstrate a method for rigorously quantifying high-dimensional entanglement from extremely limited data.  We improve an entropic, quantitative entanglement witness to operate directly on compressed experimental data acquired via an adaptive, multilevel sampling procedure. Only $6,456$ measurements are needed to certify an entanglement-of-formation of $7.11 \pm .04$ ebits shared by two spatially-entangled photons. With a Hilbert space exceeding 68 billion dimensions, we need $20$-million-times fewer measurements than the uncompressed approach and $10^{18}$-times fewer measurements than tomography. Our technique offers a universal method for quantifying entanglement in any large quantum system shared by two parties. \end{abstract}

\section*{Introduction}

Achieving a quantum advantage for information processing requires scaling quantum systems to sizes that can provide significant quantum resources, including entanglement. Large quantum systems are now realized across many platforms, including atomic simulators beyond $50$ qubits\cite{bernien:2017, zhang:2017, friis:2018}, nascent superconducting and trapped-ion based quantum computers\cite{debnath:2016, brown:2016}, integrated-photonic circuits\cite{kues:2017, carolan:2015, masada:2015, wang:2018, mennea:2018}, and photon-pairs entangled in high-dimensional variables\cite{yokoyama:2013, mirhosseini:2015, zhong:2015, xie:2015, bolduc:2016,islame:2017}.

As quantum-information-based technologies mature, it will become useful to separate the physical layer providing quantum resources (e.g trapped ions, photons) from the logical layer that utilizes those resources. For example, many imperfect qubits may form one logical qubit\cite{gambetta:2017, frowis:2017}, or thousands of atoms may coherently act as a single-photon quantum memory\cite{mcconnell:2015, tiranov:2017}. As with classical communication and computing, protocols and algorithms will be implemented in the logical layer with minimal concern for the underlying platform. Because real-world systems are varied and imperfect, the quantum resources they provide must be characterized before use\cite{gambetta:2017}.

Certifying an amount of entanglement in a large quantum system is an essential but daunting task. While entanglement witnesses\cite{terhal:2002, guhne:2009} and Bell tests\cite{brunner:2014} can reveal entanglement's presence, quantification generally requires a full estimation of the quantum state\cite{horodecki:2009}. Beyond moderately sized states, the number of parameters to physically measure (i.e. the number of the measurements) becomes overwhelming, making this approach unviable for current and future large-scale quantum technologies.

Any practical method for quantitative entanglement certification must require only limited data. Two ideas can dramatically reduce the needed measurement resources. First is the development of quantitative entanglement witnesses, which bound the amount of entanglement without full state estimation\cite{horodecki:1999,audenaert:2006, brandao:2005, eisert:2007}. In a recent landmark experiment, $4.1$ entangled bits (ebits) of high-dimensional biphoton entanglement was certified using partial state estimation\cite{martin:2017}. One ebit describes the amount of entanglement in a maximally entangled, two-qubit state\cite{horodecki:2009}. 

Second, prior knowledge can be exploited to economize sampling. Certain features, or structure, are expected in specific systems. In highly-entangled quantum systems, for example, some observables should be highly correlated, the density matrix will be low-rank, or the state may be nearly pure. Such assumptions can be paired with numerical optimization to recover signals sampled below the Nyquist limit. One popular technique is Compressed Sensing\cite{donoho:2006}, which has massively disrupted conventional thinking about sampling. Applied to quantum systems, compressed sensing reduced measurement resources significantly for tasks including tomography\cite{gross:2010, flammia:2012, tonolini:2014, kalev:2015, riofrio:2016, steffens:2017, bolduc:2017} and witnessing entanglement\cite{howland:2013, howland:2016}.

Computational recovery techniques have substantial downsides. Because they are estimation techniques, conclusions drawn from their results are contingent on the veracity of the initial assumptions. They are therefore unsuitable for closing loopholes or verifying security. Numerical solvers are often proven correct under limited noise models and require hand-tuned parameters, potentially adding artifacts and complicating error analysis. Finally, the computational resources needed become prohibitive in very large systems. The largest quantum systems characterized using these approaches remain considerably smaller than state-of-the-art. 

Here we provide an approach to entanglement quantification that overcomes these downsides. First, we improve an entropic, quantitative entanglement witness to operate on arbitrarily downsampled data. Then we develop an adaptive, multilevel sampling procedure to rapidly obtain compressed distributions suitable for the witness. Crucially, our sampling assumptions are independent of the entanglement certification, so our method can guarantee security. Because we avoid numerical optimization, error analysis is straightforward and few computational resources are needed.

\section*{Results}

\subsection{Entropic witnesses of high-dimensional entanglement}

Entanglement is revealed when subsystems of a quantum state are specially correlated. A common situation divides a system between two parties, Alice and Bob, who make local measurements on their portion. Given two mutually unbiased, continuous observables $\mbf{\hat{x}}$ and $\mbf{\hat{k}}$, they can measure discrete joint probability distributions $P(\mbf{X}_\mrm{a},\mbf{X}_\mrm{b})$ and $P(\mbf{K}_\mrm{a}, \mbf{K}_\mrm{b})$ by discretizing to pixel sizes $\Delta_{\mrm{X}}$ and $\Delta_{\mrm{K}}$. Here, bold notation indicates that $\mbf{X}$ and $\mbf{K}$ may (though need not) represent multidimensional coordinates. For example $\mbf{X}$ and $\mbf{K}$ might represent cartesian position and momentum that can be decomposed into horizontal and vertical components such that $\mbf{X}=(X,Y)$ and $\mbf{K}=(K^{(\mrm{x})},K^{(\mrm{y})})$.

A recent, quantitative entanglement witness\cite{schneeloch:2017} uses these distributions to certify an amount of entanglement:
\begin{equation}
d\log_2\left(\frac{2\pi}{\Delta_{\mrm{X}} \Delta_{\mrm{K}}}\right)-H(\mbf{X}_{\mrm{a}}|\mbf{X}_{\mrm{b}})-H(\mbf{K}_{\mrm{a}}|\mbf{K}_{\mrm{b}}) \le E_{\mrm{f}},
\label{eq:witness}
\end{equation}
where, for example, $H(\mbf{A}|\mbf{B})$ is the conditional Shannon entropy for $P(\mbf{A},\mbf{B})$. $E_{\mrm{f}}$ is the Entanglement of Formation, a measure describing the average number of Bell pairs required to synthesize the state. Eq. \ref{eq:witness} does not require full state estimation, but depends on an informed choice of $\hat{\mbf{x}}$ and $\hat{\mbf{k}}$. Still, in large systems, measuring these joint distributions remains oppressive. For example, if $\mbf{X}_{\mrm{a}}$ has $100$ possible outcomes, determining $P(\mbf{X}_{\mrm{a}},\mbf{X}_{\mrm{b}})$ takes $100^2$ joint measurements.
`
Describing quantum uncertainty with information-theoretic quantities is increasingly popular\cite{coles:2017, schneeloch:2013}. Entropies naturally link physical and logical layers and have useful mathematical properties. In particular, many approximations to the joint distributions can only increase conditional entropy. Because Eq. \ref{eq:witness} bounds $E_\mrm{f}$ from below, any such substitution is valid. 

\subsection{Improving an entropic entanglement witnesses for use with limited data}

We use two entropic shortcuts to improve the entanglement witness. First, if the system is highly entangled, and $\hat{\mbf{x}}$ and $\hat{\mbf{k}}$ are well-chosen, the joint distributions will be highly correlated; a measurement outcome for $\mbf{X}_{\mrm{a}}$ should correlate to few outcomes for $\mbf{X}_{\mrm{b}}$. The distributions are therefore highly compressible. Consider replacing arbitrary groups of elements in $P(\mbf{X}_{\mrm{a}},\mbf{X}_{\mrm{b}})$ with their average values to form a multilevel, compressed estimate $\tilde{P}(\mbf{X}_{\mrm{a}},\mbf{X}_{\mrm{b}})$. By multilevel, we mean that the new, estimated distribution will appear as if it was sampled with varying resolution---fine detail in some regions and coarse detail in others. Because coarse-graining can not decrease conditional entropy, Equation \ref{eq:witness} remains valid for $\tilde{P}(\mbf{X}_{\mrm{a}},\mbf{X}_{\mrm{b}})$ and $\tilde{P}(\mbf{K}_{\mrm{a}},\mbf{K}_{\mrm{b}})$ (see Supplemental Material: Proof arbitrary coarse-graining cannot decrease conditional entropy). 

 Good estimates for $\tilde{P}(\mbf{X}_{\mrm{a}},\mbf{X}_\mrm{b}$) and $\tilde{P}(\mbf{K}_{\mrm{a}},\mbf{K}_\mrm{b})$ can be efficiently measured by sampling at high resolution in correlated regions and low resolution elsewhere.  Note that the original ($P$) and estimate ($\tilde{P})$) are full correlation matrices with $N$ elements, but only $M\ll N$ values measured to specify $\tilde{P}$. The witness is valid for arbitrary downsampling; it works best when the approximate and actual distributions are most similar, but can never overestimate $E_\mrm{f}$ or allow false-positives.

Second, if the observables are multi-dimensional such that they can be decomposed into $d$ marginal, component observables (e.g. horizontal and vertical components) $\hat{\mbf{x}}=(\hat{x}^{(1)},\hat{x}^{(2)},...,\hat{x}^{(d)})$ (similar for $\hat{\mbf{k}}$), the conditional entropies have the property
\begin{equation}
H(\mbf{X}_\mrm{a}|\mbf{X}_\mrm{b}) \leq \sum_i^d H(X^{(i)}_\mrm{a}|X^{(i)}_\mrm{b}),
\end{equation}
with equality when $P(\mbf{X}_\mrm{a},\mbf{X}_\mrm{b})$ is separable. If we expect nearly-separable joint-distributions, the reduced, marginal joint-distributions $P(X^{(i)}_\mrm{a},X^{(i)}_\mrm{b})$ can be separately measured but still capture nearly all of the correlations present. For example, in a two-dimensional cartesian scenario, we might separately measure horizontal correlations $P(X_\mrm{a},X_\mrm{b})$, $P(K^{\mrm{(x)}}_\mrm{a},K^{\mrm{(x)}}_\mrm{b})$ and vertical correlations $P(Y_\mrm{a},Y_\mrm{b})$, $P(K^{\mrm{(y)}}_\mrm{a},K^{\mrm{(y)}}_\mrm{b})$. For $d$-component observables, this is a $d^{\text{th}}$-power reduction in the number of measurements. Like the first shortcut, this approximation also can not overestimate $E_\mrm{f}$. 

Combining both improvements, our new quantitative entanglement witness is
\begin{align}
\label{eq:impwitness}
\sum_{i=1}^d \Biggl[ \log_2 &  \left(\frac{2\pi}{\Delta_\mrm{X}^{(i)}\Delta_\mrm{K}^{(i)}}\right)  \\ \nonumber - & \tilde{H}(X^{(i)}_\mrm{a}|X^{(i)}_\mrm{b}) - \tilde{H}(K^{(i)}_\mrm{a}|K^{(i)}_\mrm{b}) \Biggr] \le E_\mrm{f}. 
\end{align}

\subsection{Proof of concept experimental setup}

As a test experimental system, we use photon pairs entangled in their transverse-spatial degrees of freedom\cite{walborn:2010, schneeloch:2016}, where the transverse plane is perpendicular to the optic axis. Our testbed, given in Figure \ref{fig:setup}(a), creates photon pairs via spontaneous parametric downconversion (see Methods). Generated photons are positively correlated in transverse-position and anti-correlated in transverse-momentum. This state closely approximates the original form of the Einstein-Podolsky-Rosen paradox. Because position $\hat{\mbf{x}}=(\hat{x},\hat{y})$ and momentum $\hat{\mbf{k}}=(\hat{k}^{\mrm{(x)}},\hat{k}^{\mrm{(y)}})$ (where $\mbf{\hat{k}}= \mbf{\hat{p}}/\hbar$) observables are continuous, this state is very high-dimensional. 

After creation, the twin photons are separated at a beam splitter and enter identical measurement apparatuses, where a basis selection system allows for interrogating position or momentum. A digital micromirror device (DMD)---an array of individually addressable micromirrors---is placed in the output plane. By placing patterns on the signal and idler DMDs and using coincidence detection, rectangular regions of the position or momentum joint-distributions are sampled at arbitrary resolution.

\subsection{Adaptive, multi-level data acquisition}

\begin{figure*}
        \centering
        \includegraphics[width=0.9\linewidth]{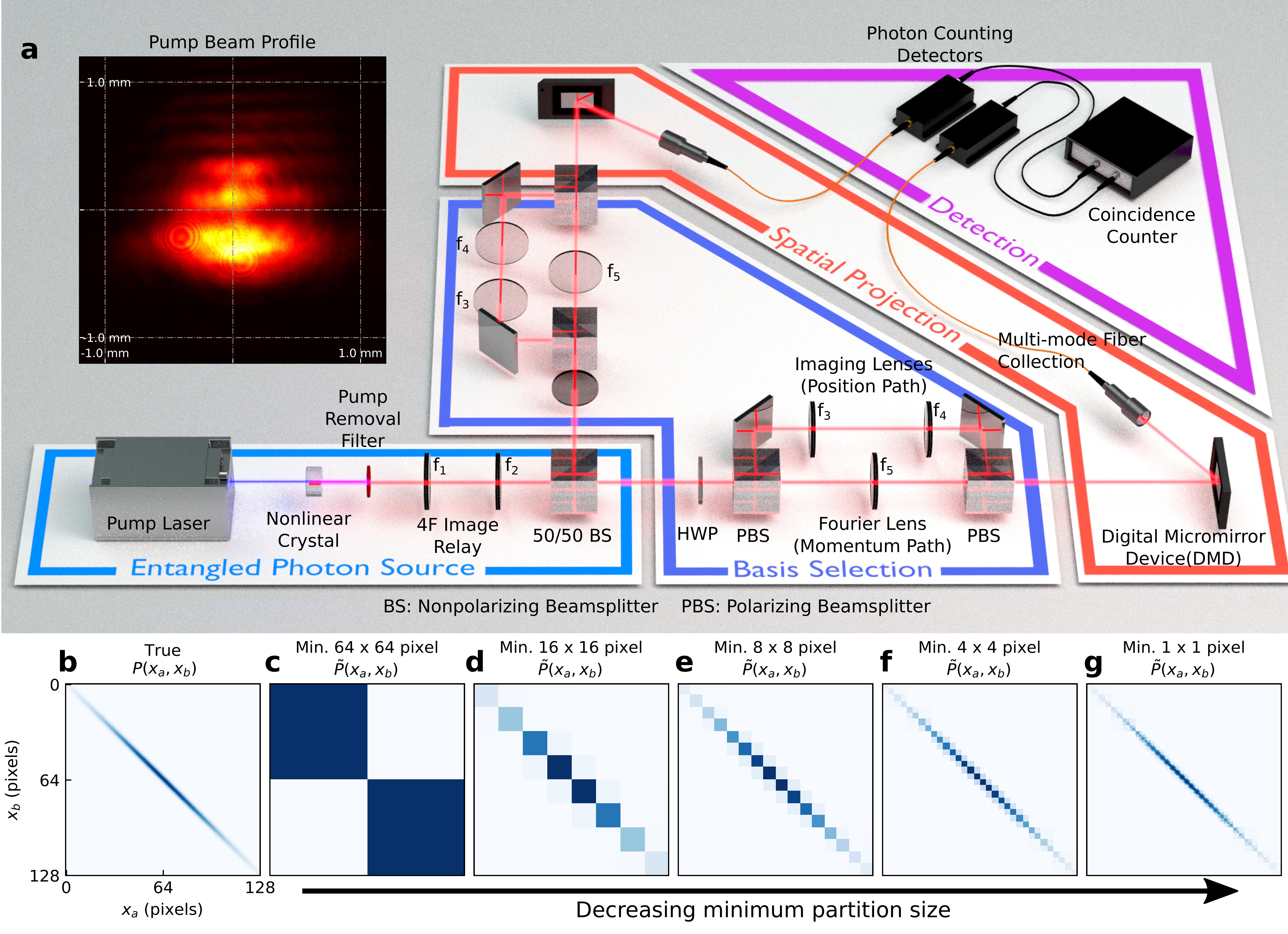}
        \caption{{\bf Experimental setup for adaptive measurements} (a) An entangled photon source produces spatially entangled photon pairs, which are separated and routed through basis selection optics that switch between measuring transverse-position or transverse-momentum. Computer-controlled digital micromirror devices and photon-counting detectors perform joint spatial projections at up to $512\times 512$ pixel resolution. (b) shows a simulated, true position joint-distribution of $P(X_\mrm{a},X_\mrm{b})$ at $128\times 128$ pixel resolution, while (c-g) show its simulated, adaptively decomposed estimate $\tilde{P}(X_\mrm{a}, X_\mrm{b})$ as it is refined to higher detail via quad-tree decomposition. When the joint-intensity in a block exceeds a user-defined threshold, it is split into four sub-quadrants and the process is recursively repeated, rapidly partitioning the space to obtain a compressed distribution from very few measurements.} 
        \label{fig:setup}
\end{figure*}

We measure joint-distributions $\tilde{P}(X_\mrm{a},X_\mrm{b})$, $\tilde{P}(Y_\mrm{a},Y_\mrm{b})$,
$\tilde{P}(K^{\mrm{(x)}}_\mrm{a},K^{\mrm{(x)}}_\mrm{b})$, and $\tilde{P}(K^{\mrm{(y)}}_a,K^{\mrm{(y)}}_\mrm{b})$. Finding compressed distributions requires a multilevel partitioning of the joint space that is not known a priori. Our adaptive approach is inspired by quad-tree image compression\cite{samet:1985}. An example is shown in Figure \ref{fig:setup}(b-g). First, all DMD mirrors are directed towards the detector to obtain a total coincidence rate $R_\mrm{T}$. Then, the joint space is divided into four quadrants (c), which are independently sampled. If the count rate in the $i^{\text{th}}$ quadrant exceeds a threshold $\alpha R_\mrm{T}$ ($0\le\alpha\le1$), the region is recursively split and the process is repeated. The algorithm rapidly identifies important regions of the joint-space for high-resolution sampling.

\begin{figure*}
        \centering
        \includegraphics[width=0.9\linewidth]{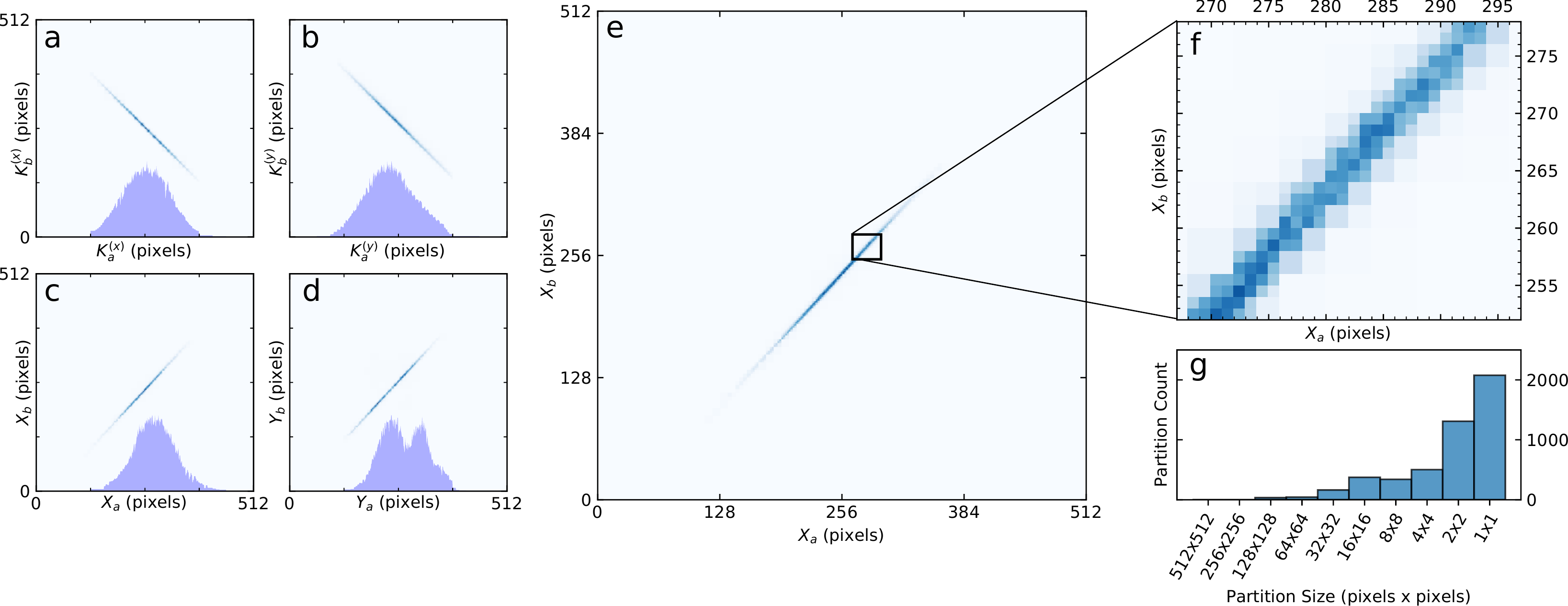}
        \caption{{\bf Measured joint probability distributions at $512\times 512$ pixel resolution}. (a-d) show the four estimated joint probability distributions with their single-party marginal distributions overlaid, showing tight correlations. (e) shows an enlarged version of $\tilde{P}(X_\mrm{a},X_\mrm{b})$ overlaid with the adaptive partitioning, with (f) showing a small central region to see fine detail. The histogram (g) shows the number of partitions as a function of their area. Only $6,456$ measurements are needed instead of $2\times 512^4$.} 
        \label{fig:distributions}
\end{figure*}

We set the maximum resolution of our system to $512\times512$ pixels-per-photon for a $512^4$-dimensional joint space. The recovered joint-distributions in position and momentum are given in Figure \ref{fig:distributions}(a-d). Figure \ref{fig:distributions}(e-f) show $\tilde{P}(X_\mrm{a},X_\mrm{b})$ with the partitioning overlaid. These display the expected strong position and momentum correlations. A histogram showing the number of partitions at various scales is given in Figure \ref{fig:distributions}(g); most partitions are either $1\times 1$ or $2\times 2$ pixels in size. Only $6,456$ partitions are needed to accurately cover the $512^4$-dimensional space---an astonishing $20$-million-fold improvement versus using the unimproved witness. Over $10^{21}$ measurements are needed to perform full, unbiased tomography.

\begin{figure*}
        \centering
        \includegraphics[width=0.7\linewidth]{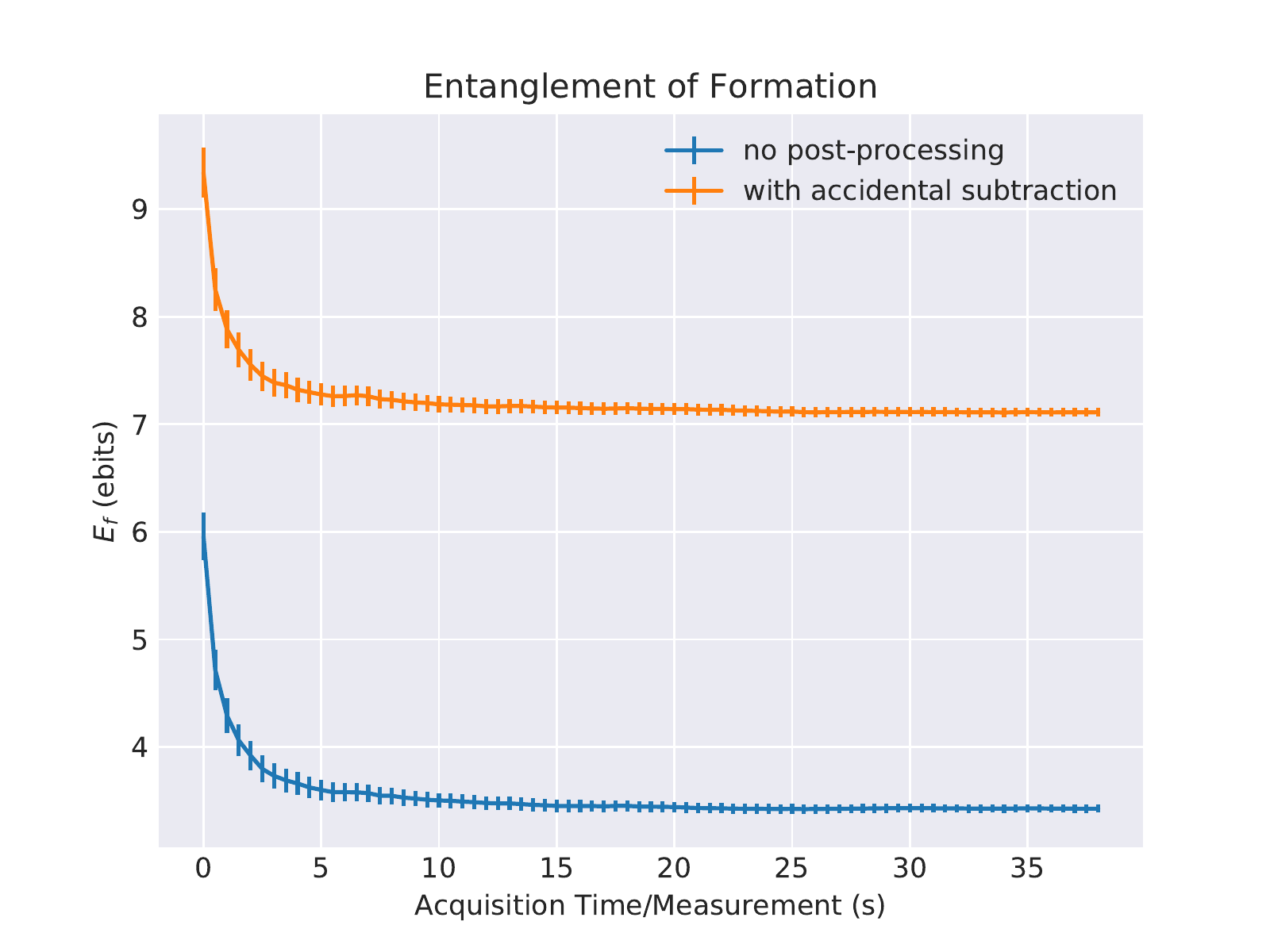}
        \caption{{\bf Entanglement quantification versus acquisition time} The entanglement of formation $E_\mrm{f}$ is given as a function of acquisition time-per-partition for unaltered coincidence data and accidental-subtracted data. Error bars enclosing two standard deviations are determined by propagation of error from photon-counting statistics. We confirm the validity of this error analysis strategy via Monte Carlo simulation in Supplemental Material: Monte Carlo error analysis (see Supplemental Figure 1).}  
        \label{fig:entropies}
\end{figure*}

The entanglement witness (Equation \ref{eq:impwitness}) applied to the data in Figure \ref{fig:distributions} is shown in Figure \ref{fig:entropies}. For short acquisition times, there is a systematic bias towards estimating a large $E_\mrm{f}$. This occurs because many of the poorly correlated regions have not yet accumulated any detection events, resulting in a systematic bias towards low conditional entropies. Statistical error is low in this region because the highly-correlated regions have high-count rates and rapidly reach statistical significance. With additional measurement time, the initial bias diminishes and statistical error decreases. To our knowledge, $7.11\pm.04$ ebits is the largest quantity of entanglement experimentally certified in a quantum system. More than $14$ maximally-pairwise-entangled logical qubits are needed to describe an equal amount of entanglement. We do not require advanced post-processing such as numerical optimization, estimation, or noise reduction; however, we do post-select on coincident detection events and optionally subtract accidental coincidences (see Methods). Our witness does not explicitly require any post-processing, and is suitable for use in adversarial scenarios given a pristine experimental system.

\begin{figure*}
        \centering
        \includegraphics[width=0.7\linewidth]{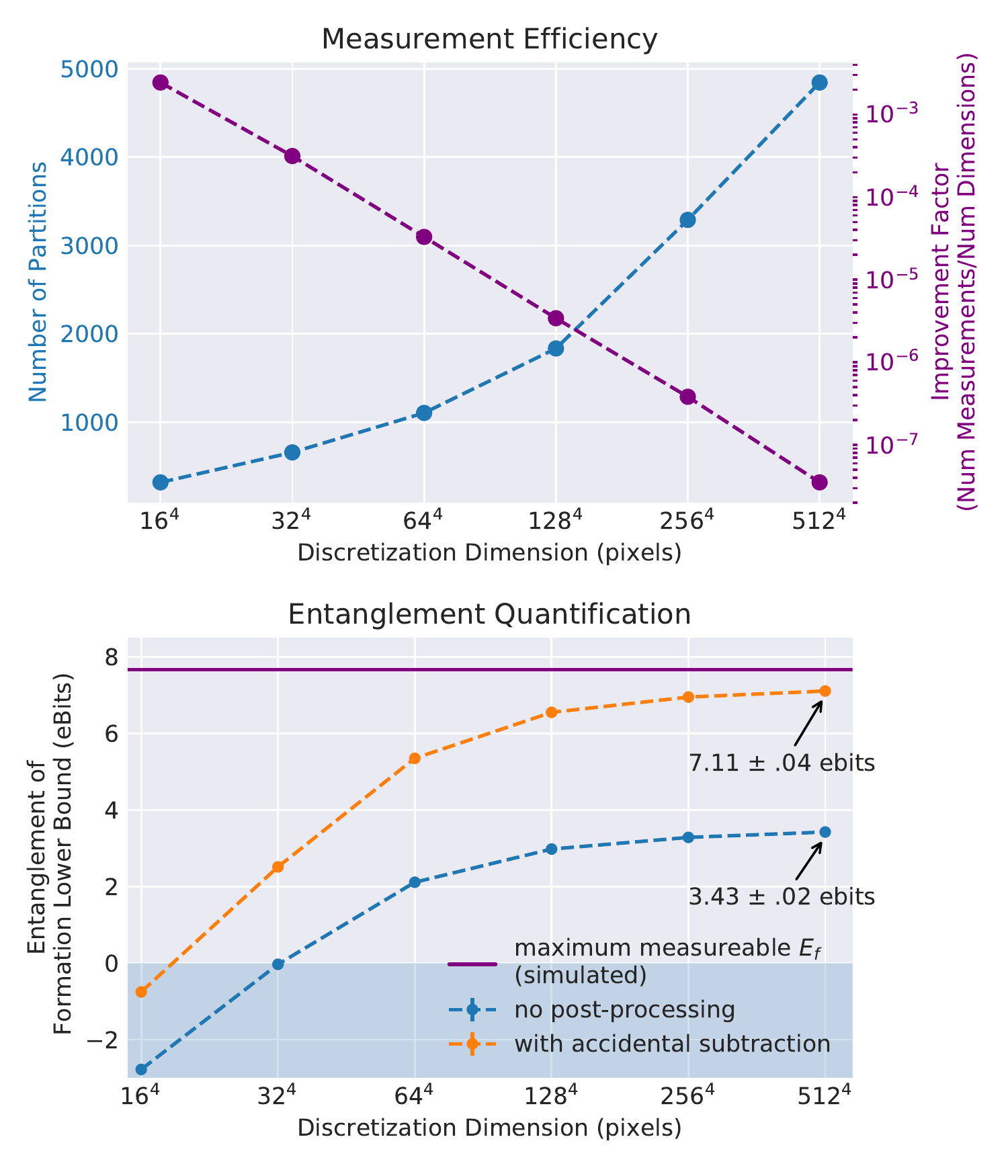}
        \caption{{\bf Entanglement quantification versus maximum resolution} (a) shows the number of partitions required as a function of maximum allowed resolution and the improvement over the uncompressed approach. (b) shows the amount of entanglement captured as the maximum resolution increases. We see the progressive nature of the technique, which witnesses entanglement with few measurements at low resolution but more accurately quantifies it with further refinement. Our results approach the ideal maximum measurable value $E_\mrm{f}=7.68$ ebits for our source.} 
        \label{fig:results}
\end{figure*}

The performance of our technique as a function of maximum discretization resolution is shown in Figure \ref{fig:results}. Figure \ref{fig:results}(a) shows the approximate distribution partition number as a function of discretization dimension and the improvement factor over naive sampling. Figure \ref{fig:results}(b) shows the certified $E_\mrm{f}$, with and without accidental subtraction, along with the ideal $E_\mrm{f}$ for our source under a double-Gaussian approximation\cite{schneeloch:2016}. Because our pump laser is not Gaussian (Figure \ref{fig:setup}(a)), the actual $E_\mrm{f}$ is slightly less but difficult to simulate. Error bars enclosing two standard deviations are scarcely visible. For low resolution, fewer than $1,000$ measurements witness entanglement. Progressively refining to higher effective resolution allows more entanglement to be certified until the maximum is reached. 

\section*{Discussion}

We have shown an efficient method for performing information-based entanglement certification in a very large quantum system. An alternative, important metric for quantifying entanglement in high-dimensional systems is the entanglement dimensionality, or Schmidt rank, which describes the number of modes over which the entanglement is distributed \cite{terhal:2000,guhne:2009, sperling:2011, krenn:2014}. In contrast, entanglement measures quantify entanglement as a resource of entangled bits without regard for their distribution. Efficiently certifying the entanglement dimensionality faces many of the same problems as certifying a number ebits, such as the intractability of full tomography and the desire to avoid side effects from prior assumptions. Recently, Bavaresco et. al. used measurements in only two bases to efficiently certify over $9$ entangled dimensions between orbital-angular-momentum entangled photon pairs without special assumptions about the underlying state \cite{bavaresco:2018}. 

The number of entangled dimensions and the number of entangled bits are complementary but distinct characterizations of entanglement \cite{erker:2017}. If a density matrix can not be decomposed into pure states with Schmidt rank less than $d$, then the state is at least $d$-dimensionally entangled. However, a $d$-dimensional entangled state may possess an arbitrarily small amount of entanglement. Consider a system with a large Schmidt rank, but where one coefficient of the Schmidt decomposition is much larger than the others. This system will have a large entanglement dimensionality but require few entangled bits to synthesize. In this way, a given entanglement dimensionality $D$ provides an upper bound on the entanglement of formation $E_\mrm{f}$ such that $0<E_\mrm{f}\le \log_2 D$. In contrast, a given $E_\mrm{f}$ provides a lower bound to the entanglement dimensionality $D \ge 2^{E_\mrm{f}}$, describing the situation where all $D$ dimensions are maximally entangled. Our quantitative witness therefore also certifies entanglement dimensionality, but may dramatically underestimate when the target system is not near-maximally entangled (e.g. with additive noise or non-uniform marginals). In our case, we certify $2^{7.11}\ge138$ maximally-entangled dimensions with background subtraction and $2^{3.43}\ge10$ maximally-entangled dimensions without background subtraction. To our knowledge, $10$ entangled dimensions is the largest certified entanglement dimensionality without assumptions about the state.

Our approach shows a path forward for certifying quantum resources in large quantum systems, where we exploit prior knowledge without conventional downsides. We show the power of an information-theoretic approach to characterizing quantum systems, and how compression can be leveraged without computational signal recovery. Though the method presented here is limited to EPR-style systems where entanglement is shared by two parties, we expect similar techniques for many-body systems utilizing higher-order correlations will soon follow.

\begin{methods}

\subsection{Experimental apparatus}
$810$ nm, spatially entangled photon pairs are produced via spontaneous parametric downconversion (SPDC)\cite{schneeloch:2016}. The pump laser is a $405$ nm diode laser (CrystaLaser DL405-025-SO) attenuated to $7.9$ mW with a $356$ $\mu$m (x) $\times$ $334$ $\mu$m (y) beam waist. A spectral clean-up filter (Semrock Versachrome TBP01-400/16) removes unwanted $810$ nm light. The pump laser is not spatially filtered. The nonlinear crystal is a $3$ mm long BiBO crystal oriented for type-I, degenerate, collinear SPDC. The crystal is held at $32.3^{\circ}$C in an oven for long-term stability. A low-pass interference filter (Semrock LP442) removes remaining pump light, followed by a telescope relay system ($f_1=50$ mm, $f_2=100$ mm) that magnifies the SPDC field $\approx 2$X. A half-waveplate and polarizing beamsplitter choose between imaging ($\hat{\mbf{x}}$) and Fourier-transforming ($\hat{\mbf{k}}$) beam-paths; a beam block is placed in the unused path. 

The DMDs (TI Lightcrafter 4500) are computer controlled via a digital video port (HDMI). A $512\times1024$ physical-pixel area was used for data given in this manuscript. Because the DMD has twice the vertical pixel density, this corresponds to a square area. $10$ mm effective focal length, aspheric lenses (Thorlabs AC080-010) couple light into $100$ micron core multi-mode fibers connected to photon-counting detector modules (Excelitas SPCM-AQ4C-10). $810/10$ nm bandpass filters (Thorlabs FBS810-10) are placed before the fiber coupling. A time-correlated single-photon counting module (PicoQuant HydraHarp400) produces histograms of photon-pair relative arrival times. We post-select on coincident detections within a $1$ ns coincidence window centered on the histogram peak. With all DMD mirrors pointed towards the detectors, there are approximately $26,400$ total coincidences/second.

\subsection{Data collection} 

The apparatus must be adjusted to separately measure the four reduced, joint-probabilty distributions $P(X_\mrm{a},X_\mrm{b})$, $P(Y_\mrm{a},Y_\mrm{b})$, $P(K^{\mrm{(x)}}_\mrm{a}, K^{\mrm{(x)}}_\mrm{b})$, and $P(K^{\mrm{(y)}}_\mrm{a},K^{\mrm{(y)}}_\mrm{b}).$ For example, to access the horizontal, joint-position distribution $P(X_\mrm{a}, X_\mrm{b})$, we adjust the half-waveplates to direct light down the imaging beam-paths so the DMDs lie in an image plane of the nonlinear crystal. To access a particular, rectangular element of the distribution, local, one-dimensional "top-hat" patterns are placed on signal ($\mathrm{a}$) and idler ($\mathrm{b}$) DMDs that only vary horizontally. In the regions where light should be directed to the detectors, all vertical pixels are used. The local images' outer-product defines the rectangular region of the joint-space $P(X_\mrm{a},X_\mrm{b})$ that is being sampled. 

To instead access the vertical, joint-position distribution $P(Y_\mrm{a},Y_\mrm{b})$, local DMD patterns are used that only vary vertically. The joint-momentum distributions are similarly sampled, with the half-waveplates instead adjusted to send light down the Fourier transforming optical path so that the DMDs sit in the far-field of the nonlinear crystal. 

\subsection{Adaptive Sampling Algorithm}

For each configuration, experimental data is stored in nodes in a quad-tree decomposition of $P$ whose levels describe increasingly fine detail. The $i^{\mrm{th}}$ node corresponds to a square area of $\tilde{P}$ at location $(x^{i}_\mrm{a},x^{i}_\mrm{b})$ with span $w^{i}_\mrm{a}=w^{i}_\mrm{b}=w$. Nodes are sampled by placing the corresponding, one-dimensional local patterns on the DMDs and generating a coincidence histogram during acquisition time $T_\mrm{a}=0.5$ s. Coincidences $C_i$ are counted within a $1$ ns coincidence window centered on the coincidence peak; accidental coincidences $A_i$ are counted in a $1$ ns window displaced $2$ ns from the coincidence window. Coincidence and accidental values are appended to a list each time the node is sampled. The estimated count-rate $R_i=\braket{C_i}/\epsilon_i T_\mrm{a}$, where $\epsilon_i$ is a calibrated, relative fiber coupling efficiency. Optionally, $A_i$ can be subtracted from $C_i$ for accidental removal. Uncertainty is computed by assuming Poissonian counting statistics for $C_i$ and $A_i$ and applying standard, algebraic propagation of error through the calculation of the entanglement quantity (Eq. \ref{eq:impwitness}). 

The data collection algorithm consists of a partitioning phase followed by an iterative phase. During partitioning, the algorithm repeatedly iterates through a scan-list of leaves of the tree. Node $i$ is considered stable when $\mathrm{sgn}(\alpha R_\mrm{T}-R_i)$ is known to at-least $\beta$ standard-deviations of certainty, where splitting threshold $\alpha$ ($0 \le \alpha \le 1$) and stability criterion $\beta$ are user-chosen heuristics. Stable nodes are no longer measured. If a node is stable and $R_i \ge \alpha R_\mrm{T}$, the node is split into four equal-sized sub-quadrants which are initially unstable and added to the scan-list. Optionally, a maximum resolution  (maximum tree depth) may be set.

The transition to the iterative phase occurs when the percentage of unstable leaves is less than $\Gamma$, a user chosen parameter. At this point, stability is ignored and all leaf nodes are scanned repeatedly and guaranteed to have the same total acquisition time. Various final stopping criteria can be used; we chose a fixed total run time. Note that heuristic parameters $\alpha$, $\beta$, and $\gamma$ may be changed during operation if desired. For the data shown in this manuscript, $\alpha=.002$, $\beta=2$, and $\Gamma=.15$ with a $30$ hour runtime.

The probability distribution $\tilde{P}$ is computed by uniformly distributing the estimated count rate (with or without-accidental subtraction) from each leaf node across its constituent elements in $\tilde{P}$, followed by normalization.

\section*{Data Availability}
The data supporting the results presented in this manuscript is available from the corresponding author G.A.H upon request.

\end{methods}


\clearpage

\begin{addendum} 
\item[Acknowledgements] We gratefully acknowledge support from the OSD ARAP QSEP program and Air Force Office of Scientific Research LRIR 14RI02COR. J.S. acknowledges support from the National Research Council Research Associate Program. Any opinions, findings, and conclusions or recommendations expressed in this article are those of the authors and do not necessarily reflect the views of AFRL.

\item[Author Contributions] G.A.H. and J.S. conceived of the idea and contributed equally. J.S. derived the entanglement witness and led the theoretical analysis.  G.A.H and C.C.T. developed the data collection algorithm. G.A.H performed the experiment with help from M.L.F and analyzed the data with help from C.C.T. and J.S. P.M.A. participated in useful scientific discussions. G.A.H wrote the manuscript with contributions from all authors. 

\item[Competing Interests] The authors declare that they have no
competing interests.

\item[Correspondence] Correspondence and requests for materials
should be addressed to G.A.H.~(email: gaheen@rit.edu).

\end{addendum}
\bibliography{refs}

\clearpage
\section*{Supplemental Material}
\setcounter{figure}{0}

\input{supplemental.tex}
\end{document}

%% file: supplemental.tex
\renewcommand{\figurename}{Supplemental Figure}







\subsection{Proof that arbitrary coarse-graining cannot decrease conditional entropy}

We are given two discrete probability distributions $P_{1}(X_{\mrm{A}},X_{\mrm{B}})$ and $P_{2}(X_{\mrm{A}},X_{\mrm{B}})$. We will also assume a permutation operation $\chi$, that shuffles the outcomes of $X_\mrm{A}$ and $X_\mrm{B}$. With this, we define the permuted distributions $P_{1}^{'}(X_\mrm{A},X_\mrm{B})$ and $P_{2}^{'}(X_\mrm{A},X_\mrm{B})$ as the result of permutation operator $\chi$ on  $P_{1}(X_\mrm{A},X_\mrm{B})$ and $P_{2}(X_\mrm{A},X_\mrm{B})$, respectively. 

The joint convexity of relative entropy states that given distributions $P_{1}, P_{2}, P_{1}^{'}$, and $P_{2}^{'}$, the following inequality holds:
\begin{align}
\label{ConvRelEnt}
\lambda \mathscr{D}&(P_{1}||P_{2}) + (1-\lambda)\mathscr{D}(P_{1}^{'}||P_{2}^{'})\geq \nonumber\\
&\geq \mathscr{D}(\lambda P_{1} + (1-\lambda) P_{1}^{'}||\lambda P_{2} + (1-\lambda) P_{2}^{'})
\end{align}
where $\lambda\in[0,1]$.

Next, we define the mixed probability distribution $\bar{P}_{1}\equiv(\lambda P_{1} + (1-\lambda) P_{1}^{'})$, and define $\bar{P}_{2}$ similarly. Since $P_{1}^{'}$ and $P_{2}^{'}$ are respectively related to $P_{1}$ and $P_{2}$ by the same permutation $\chi$, we have that $\mathscr{D}(P_{1}||P_{2})=\mathscr{D}(P_{1}^{'}||P_{2}^{'})$. Therefore, we obtain the inequality:
\begin{equation}\label{RelEntMixIneq}
\mathscr{D}(P_{1}||P_{2})\geq \mathscr{D}(\bar{P}_{1}||\bar{P}_{2}).
\end{equation}
This result that mixing (i.e., majorization) cannot increase relative entropy has far reaching applications. In particular, coarse-graining is a form of majorization between adjacent elements in a probability distribution. Because all (Shannon) entropic functions can be expressed in terms of relative entropies, it immediately follows that:
\begin{align}
H_{\mrm{\bar{P}}}(X_\mrm{A})&\geq H_{\mrm{P}}(X_\mrm{A})\\
H_{\mrm{\bar{P}}}(X_\mrm{A},X_\mrm{B})&\geq H_{\mrm{P}}(X_\mrm{A},X_\mrm{B})\\
H_{\mrm{\bar{P}}}(X_\mrm{A}|X_\mrm{B})&\geq H_{\mrm{P}}(X_\mrm{A}|X_\mrm{B})
\end{align}
where the subscripts $\mrm{P}$ and $\mrm{\bar{P}}$ represent the probability distribution before and after coarse-graining, respectively. In addition, the mutual information and the conditional mutual information obey the inequalities
\begin{align}
H_{\mrm{\bar{P}}}(X_\mrm{A}:X_\mrm{B})&\leq H_{\mrm{P}}(X_\mrm{A}:X_\mrm{B})\\
H_{\mrm{\bar{P}}}(X_\mrm{A}:X_\mrm{B}|X_\mrm{C})&\leq H_{\mrm{P}}(X_\mrm{A}:X_\mrm{B}|X_\mrm{C}).
\end{align}
where again, the subscripts $\mrm{P}$ and $\mrm{\bar{P}}$ denote the true and coarse grained probability distribution, respectively. Furthermore, both the continuous mutual information $h(x_\mrm{A}:x_\mrm{B})$ and the continuous conditional mutual information $h(x_\mrm{A}:x_\mrm{B}|x_\mrm{C})$ are expressible as high-resolution limits of corresponding discrete mutual informations. Because successive coarse grainings cannot increase these quantities, the following inequalities hold between discrete and continuous mutual information
\begin{align}
h(x_\mrm{A}:x_\mrm{B})&\geq H(X_\mrm{A}:X_\mrm{B})\label{mineq}\\
h(x_\mrm{A}:x_\mrm{B}|x_\mrm{C})&\geq H(X_\mrm{A}:X_\mrm{B}|X_\mrm{C})\label{newineq}
\end{align}
While the former inequality \eqref{mineq} can be found with alternative methods, the latter inequality \eqref{newineq} is new to the literature.

\subsection{Proof of inequality 2}

Inequality (2) derives from two fundamental properties of Shannon entropy. To expand notation, we have:
\begin{equation}
H(\mathbf{X}_\mrm{a}|\mathbf{X}_\mrm{b})\equiv H(X_\mrm{a}^{(1)},...,X_\mrm{a}^{(d)}|X_\mrm{b}^{(1)},...,X_\mrm{b}^{(d)})
\end{equation}
First, is that the joint Shannon entropy is less than or equal to the sum of the marginal entropies:
\begin{equation}
H(\mathbf{X}_\mrm{a}|\mathbf{X}_\mrm{b})\leq\sum_{i=1}^{d}H(X_\mrm{a}^{(i)}|X_\mrm{b}^{(1)},...,X_\mrm{b}^{(d)})
\end{equation}
Second, is that conditioning on additional variables cannot increase entropy, or conversely that removing conditioning variables cannot reduce entropy:
\begin{equation}
H(X_\mrm{a}^{(i)}|X_\mrm{b}^{(1)},...,X_\mrm{b}^{(d)})\leq H(X_\mrm{a}^{(i)}|X_\mrm{b}^{(i)})
\end{equation}
Together, this proves inequality (2):
\begin{equation}
H(\mathbf{X}_\mrm{a}|\mathbf{X}_\mrm{b})\leq \sum_{i=1}^{d}H(X_\mrm{a}^{(i)}|X_\mrm{b}^{(i)}).
\end{equation}

\subsection{Monte Carlo error analysis}

For the results shown in the manuscript, we used standard, first-order propagation-of-uncertainty for error analysis. Each coincidence-count measurement is assumed to have Poissonian uncertainty, and this uncertainty is analytically propagated through the analysis (e.g. $f(x_0\pm \delta) = f(x_0)  \pm \left (\frac{df}{dx}\right)_{x_0} \delta$).

To confirm the validity of our propagation-style error analysis, we also estimated our uncertainty with Monte Carlo simulations. This approach does not suffer any potential issues that may arise where our equations may not be sufficiently well-behaved for the first-order propagation of error. However, it does replace a simple analytical result with the need for computational simulations.

To perform Monte Carlo simulations, each coincidence count measurement is used to sample from a Poissonian distribution. Then, we follow our previously described process for generating joint-probability distributions (with or without accidental subtraction) and calculating the amount of entanglement. This process is repeated many times to see how the Poissonian counting statistics propagate to our final result.

In Supplemental Figure \ref{fig:supp:mc}, we recreate Figure 3 from the main text using this approach with $100$ trials. The error bars shown enclose two standard deviations. The uncertainties from this approach behave similarly to the analytic propagation-of-error used in the main manuscript, however the uncertainties are even smaller. The values obtained for the entanglement of formation are $7.154 \pm .015$ $(7.112 \pm .0412)$ ebits with background subtraction and $3.459 \pm .012$ $(3.425 \pm .038)$ ebits, where the analytic result is given in parentheses. The two outcomes are in good agreement, with between two-times and four-times lower uncertainty with the Monte Carlo simulations.

\subsection{Maximum possible entanglement that can be certified with this technique}

For photon statistics contained within a finite window, the maximum possible entanglement our relation can characterize is when a pixel in the signal arm is correlated to only a single pixel in the idler arm, or when all conditional entropies are zero. In this case, the inequality reads:

\begin{equation}
E_{\mrm{f}}\geq \log\Bigg(\frac{(2\pi)^{2}}{\Delta x_{\mrm{A}} \Delta y_{\mrm{A}} \Delta k_{\mrm{xA}}\Delta k_{\mrm{yA}}}\Bigg).
\end{equation}

For perfect diagonal correlations, the number of measurements we need with our technique scales favorably with resolution, improving better with tighter correlations. For example, for $N\times N$ resolution in both position and momentum (assuming $N$ is a power of two for simplicity), then one needs only about $12(N-\log_{2}(N) -2)$ measurements, which, for $N=512$ would be about 6096 measurements. This does not include the number of measurements needed to acquire this partitioning, which scales similarly. When the correlations are less tight, more pixels are required at maximum resolution, increasing this total.

\begin{figure*}
        \centering
        \includegraphics[width=0.7\linewidth]{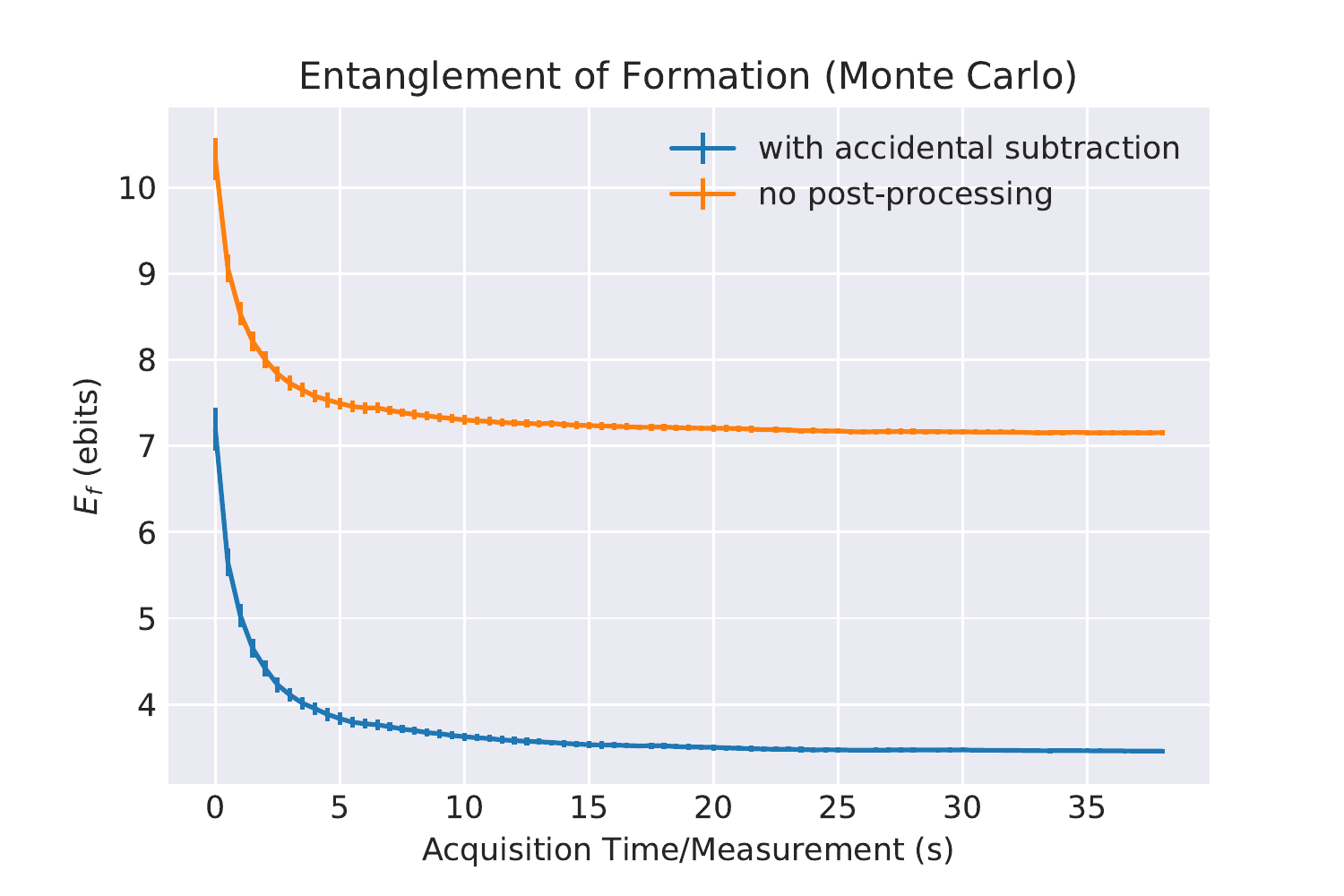}
        \caption{{\bf Entanglement quantification versus acquisition time with Monte Carlo uncertainty analysis} Measured coincidence counts are used to draw values from a Poisson distribution for 100 trials. Error bars enclose two standard deviations and are in good agreement with the analytical approach to error analysis used the main text (see Figure 3).} 
        \label{fig:supp:mc}
\end{figure*}
